\documentclass[11pt,a4paper]{article}
\pdfoutput=1
\usepackage{graphicx}
\usepackage{amsmath}
\usepackage{mathptmx}	% times for both maths and rm

\newcommand\etal{\mbox{\textit{et al.}}}
\title{Predicting structural and statistical features of wall turbulence}
\author{B.~J.~McKeon$^{1}$, A.~S.~Sharma$^2$ and I. Jacobi$^{1}$}

\begin{document}
\maketitle

\begin{enumerate}
 \item Graduate Aerospace Laboratories, California Institute of Technology, U.S.A.
 \item Department of Aeronautics, Imperial College, London, U.K.
\end{enumerate}

\begin{abstract}
% This is the most important paragraph in the whole paper
\emph{The majority of practical flows, particularly those flows in applications of importance to transport, distribution and climate, are turbulent and as a result experience complex three-dimensional motion with increased drag compared with the smoother, laminar condition. In this study, we describe the development of a simple model that predicts important structural and scaling features of wall turbulence. We show that a simple linear superposition of modes derived from a forcing-response analysis of the Navier-Stokes equations can be used to reconcile certain key statistical and structural descriptions of wall turbulence. The computationally cheap approach explains and predicts vortical structures and velocity statistics of turbulent flows that have previously been identified only in experiments or by direct numerical simulation. In particular, we propose an economical explanation for the meandering appearance of very large scale motions observed in turbulent pipe flow, and likewise demonstrate that hairpin vortices are predicted by the model.
This new capability has clear implications for modeling, simulation and control of a ubiquitous class of wall flows.}
\end{abstract}

%[One or two sentences providing a basic introduction to the field, comprehensible to a scientist in any discipline.]
In flows of the most widely-ranging practical interest, namely flows over surfaces or \emph{wall turbulence}, the turbulence problem is greatly complicated by inhomogeneity in the wall-normal direction caused by the no-slip and no-penetration boundary conditions at the wall.  As such, there has been an enduring practical and intellectual interest in understanding how to describe the state of a range of turbulent flows with a view to controlling and optimizing their behavior, which could help reduce drag losses in applications like transport vehicles or pipelines.

The nonlinear equations of motion for fluids were written down by Navier and Stokes in the early 19th century. However turbulence, a phenomenon involving interaction at a very broad range of spatial and time scales, has proved difficult to model without resorting to expensive direct numerical simulation.
%[Two to three sentences of more detailed background, comprehensible to scientists in related disciplines.]
Turbulent fluctuations of velocity and pressure have characteristics of chaotic motion, and a common feature of wall turbulence is the presence of persistent, coherent vortical structures which inhabit distinct regions of the flow \cite{Theodorsen52,Schoppa02} and have been generally believed to be dominated by nonlinear dynamics. These observations of distinct classes of coherent structure in parallel with the measurement of turbulent velocity statistics over the last sixty or more years have led to a dichotomy in the understanding of the formation, development and scaling of even the simplest, canonical turbulent wall flows, such as flow through straight channels and pipes, or flow over a flat plate in the absence of a pressure gradient.
%[One sentence summarising the main result (with the words “here we show” or their equivalent).]
%[Two or three sentences explaining what the main result reveals in direct comparison to what was thought to be the case previously, or how the main result adds to previous knowledge.]

%Previously, the hairpin vortices observed in experiments have lacked an \emph{ab initio} mathematical description and have been described phenomenologically and via the physical processes likely to drive them, particularly in the absence of a full velocity field \cite{Perrychong82,Zhou99}.

In this work, we demonstrate that the formation of such coherent structures is a natural consequence of the velocity field that arises as a near-singular response of the linearised equations of motion to the wave-like forcing arising due to nonlinear interactions with other wave-like motions. The framework provides predictive information both about how the wall-normal distribution of turbulent energy of the velocity field shows self-similarity across Reynolds numbers, and also describes the wall-normal coherence of structures within the flow. The predictions of the scaling and distribution of second-order velocity statistics, and of structural characteristics are provided using an essentially linear model. The analysis uses only the Navier-Stokes equations and an assumed mean velocity profile.

% A brief introduction to structures in turbulence
The existence of coherent vortical structures in wall turbulence has been known for many decades \cite{Theodorsen52}, but a quantitative or \emph{ab initio} mathematical means of describing the full flow field in terms of these building blocks has been lacking. There have been many observations of a common structural pattern described as a hairpin vortex (a vortical loop with legs originating close to the wall, a body inclined in the downstream direction and a sense of rotation consistent with the vorticity associated with the mean shear) in both experiment and simulation \cite{HB81,Adrian00,Adrian07,Wu08}, while the attached eddy hypothesis formulated by Townsend \cite{Townsend} and Perry and co-authors \cite{Perrychong82} in the mid-20th century has sought to use phenomenological arguments to determine the velocity field associated with hierarchies of these structures. Nonlinear mechanisms for the formation and growth of packets of hairpins in otherwise quiescent flow have also been proposed \cite{Zhouvort96}, while there remains controversy as to whether they are simply a remnant of the transition to turbulence, or even whether whole hairpins, rather than a statistical vortical imprint, even exist.

With alternative models lacking, statistical descriptions have remained the simplest method to obtain comprehensive descriptions of the turbulent field.  Under the assumptions of stationarity and ergodicity, simple spectral representations provide information on important scales in the flow. While information at streamwise and spanwise wavenumber and temporal frequency ($k,n,\omega$, respectively) is required to fully describe the flow, usually one-dimensional (integrated) spectra are reported due to the intensive nature of obtaining and storing information in three dimensions. The streamwise spectra reveal the presence of the widest range of scales, from the Kolmogorov scales \cite{K41,K62} responsible for small-scale dissipation of energy up to so-called \emph{very large scale motions (VLSMs)} of the order of ten times the length scale imposed by the flow geometry (typically the pipe radius, or the boundary layer thickness). Some researchers have inferred that the VLSMs reach lengths of the order of thirty radii when spanwise \textit{meander} of the coherent region is taken into account~\cite{Monty07}, a topic of continuing interest in the literature. A non-dimensional representation of ratio of the scales that are important far from the wall and in the near-wall region is given by the Reynolds number $Re$, a parameter that rapidly increases with increasing velocity.\footnote{The Reynolds number described here is defined by the ratio between a flow length scale (here the pipe radius $R$) and a scale representative of small, near-wall motions, the viscous length scale defined by $\nu/u_\tau$, where $\nu$ is the kinematic viscosity of the fluid and $u_\tau$ is a velocity scale associated with the skin friction acting on the wall, $\tau$, and the fluid density.  Thus our Reynolds number is given by $Re=R^+=Ru_\tau/\nu$ (alternative definitions are possible).}  A recent focus of research has been on the origin and development of large and small scales, and how their competing influences lead to trends with increasing Reynolds number.  While the origin of the VLSMs has remained elusive, there has been recent progress in predicting velocity statistics by considering the nature of their spectral interaction with the energetic near-wall region~\cite{Mathis09}.

A truly complete interpretation of a turbulent flow field requires assimilation of both the velocity statistics and vortical structures; observational progress has been made by considering the generation of one by the other, posing a classic ``chicken and egg'' conundrum. In the absence of low-order or simple predictive models for turbulent flows, the only current recourse for the fluid mechanician who wishes to quantify the state of turbulence lies in expensive simulations or experiments.

%\BJM{??? Remove? Turbulence frequently occurs in flows where infinitesimal perturbations to the laminar flowfield asymptotically decay. Larger perturbations however lead to transition and then turbulence. Systems that behave this way generally are considered to be essentially nonlinear.}

One recent and promising approach to understanding the development of the turbulent field has been the analysis of the receptivity of wall flows to particular forcing or disturbances. In the case of transition to turbulence from laminar flow, progress has been made \cite{Schmid} by considering the very large amplification that is possible arising from the evolution of finite perturbations to the laminar flows. This approach is best suited to explaining sub-critical transition to turbulence and it is unclear how suitable such an analysis is in fully turbulent flow. The approach generalises straightforwardly to complex geometries \cite{Abdessemed09} and the current approach should also generalise in such a way.

There has been some investigation of the non-normal growth mechanisms in turbulent flow, for instance the early attempt by \cite{Butler92} to predict the spacing of near-wall streaks in turbulent flow and the more recent studies of \cite{delAlamo06}, \cite{Cossu09} and \cite{Willis09}. These approaches exploit the properties of the linear operator to ``select'' by preferential amplification the ``optimal'' disturbance that leads to the largest amplification in some norm across all disturbances. A notable and somewhat limiting issue in the treatment of turbulent flow is the modelling of the interaction of the amplified disturbances with the ``background'' turbulence. One solution proposed \cite{Reynolds72} was to use the eddy viscosity formulation of \cite{Cess58}; this relies on an \emph{a priori} knowledge of the spatially-averaged, wall-normal variation of the mean Reynolds stress integrated across contributions from various Reynolds numbers. Other attempts have been made to use linear analysis to explain the dominant features of turbulent flow in terms of optimal transient modes in the initial value problem~\cite{Butler92, delAlamo06, Cossu09}, response to stochastic forcing~\cite{Farrell98, Bamieh01} and system norm analysis~\cite{Jovanovic05,Meseguer03}. In recent work, \cite{Willis09} have investigated the maximal response to harmonic forcing in pipe flow. Perhaps most importantly, it has been shown that both linear non-normality~\cite{Henningson94} and the terms that are linear in the turbulent fluctuation~\cite{Kim00} are required to sustain turbulence in infinite or periodic wall-bounded flows.

Alternatively it is possible to examine the response of the linearised system to forcing. In this picture, the observed behaviour is explained by even small forcing on the system leading to energetic flowfield response \cite{Bamieh01, Jovanovic05}.
The device of identifying the nonlinear interaction between Fourier modes in the Navier-Stokes equations (NSE) as a forcing actually acting on the a linear system permits the extension of such methods to fully developed turbulent flows \cite{McKeon2010}, allowing the successful prediction of observed features of turbulent flow.

%Universality and scaling are very important insights, classing high-Re turbulent flows together.
%Turbulent flow is made up of structural building-blocks, notably rolls \& streaks, hairpins, vortex rings. It has been recently found that very large %structures also seem to be important at transferring energy through the flow and modifying the expected scaling close to the wall.
%These structures are typically described in a nonlinear or qualitative way.
%Secondly, there is both old \& new understanding that linear processes are important to receptivity in turbulence.
%Thirdly, turbulent flows are often described in statistical terms as moments of stationary distributions about the local mean velocity as a function of %wall-normal distance, $U(y)$.
%Near-wall cycle responsible for near-wall peak in streamwise stress.

% phrase this sentence carefully to avoid hyperbole
%Repetition (BJM) Here we describe a simple analysis that reconciles many of the important predictions and observations of the structural, statistical and receptivity viewpoints.
We consider turbulent flow through a long straight pipe with a cylindrical cross-section. Laminar flow in this geometry is stable to infinitesimal disturbances, and the transition to turbulence is still not completely understood  \cite{HofSci10}, but the pipe offers the analytical benefits of statistical homogeneity in the streamwise direction and a simple constraint on the azimuthal wavenumber. Turbulent flow through pipes is important for applications like the transport of fluids such as oil and natural gas, and also in numerous natural and biomedical applications, and is also highly relevant to the study of other canonical flows.  % Repetition (BJM): The approach detailed here requires minimal modification to consider such other flows.  , a topic of ongoing work.

% Key assumptions
We recently formulated a traveling wave framework by which to analyze the dominant velocity mode shapes at particular wavenumber-frequency combinations \cite{McKeon2010}. The fully turbulent velocity field, $\mathbf{v}$, can be represented in a divergence-free basis as a superposition of Fourier modes at various spatial wavenumbers and temporal frequencies with wall-normal variation (waves),
\[\mathbf{v}(y,x,\theta,t)=\sum_{n}\int^{\infty}_{-\infty}\int^{\infty}_{-\infty}\mathbf{A}_{k,n,\omega}(y)e^{i(\omega t - kx -n\theta)} dk d\omega,\]
where $y$ is the wall-normal distance, $k$ and $n$ are the wavenumbers in the streamwise ($x$) and azimuthal ($\theta$) directions, all normalized with the pipe radius, and $\omega$ is the temporal frequency with respect to time $t$.
These waves, which are helical in pipe flow, represent the transmission of fluctuating energy relative to the mean flow. Under this decomposition, the Navier-Stokes equations can be written in an input-output formulation, where the characteristics of the linear operator $\mathcal{L}_{k,n,\omega}(y)$ describe the velocity field's response to an unmodelled harmonic forcing $\mathbf{f}$, itself arising from the nonlinear interactions between the velocity field at other wavenumber-frequency combinations. The equations expressed in this form are
\[\mathbf{v}(y,k,n,\omega)=(i\omega - \mathcal{L}_{k,n,\omega}(y))^{-1}\mathbf{f}(y,k,n,\omega)\]
and the operator $(i\omega - \mathcal{L}_{k,n,\omega}(y))^{-1}$ relating the forcing to the velocity field response is called the \emph{resolvent}. % or sometimes, the transfer function.
%where $\mathcal{L}_{k,n,\omega}$ has the following form
%\begin{equation}
%\mathcal{L}_{k,n,\omega}^{-1}=
%\left(
%  \begin{array}{c}
%    -{A}{Re^{-1}}-ikU + i\omega ~~~~~~~~~~ -B ~~~~~~~~~~~~~~~~~~ 0 \\
%    B ~~~~~~~~~~~~~~~~~~~~ -{A}{Re^{-1}}-ikU + i\omega ~~~~~~~~~~ 0 \\
%    \partial_r U ~~~~~~~~~~~~~~~~~~~~~~~~~~~~~~~~~~~~~~~~ 0 ~~~~~~~~~~ -{A}{Re^{-1}}-ikU + i\omega \\
%  \end{array}
%\right)^{-1}
%\label{eqn:resolvent}
%\end{equation}
%Here $A$ and $B$ are differential operators that can be simply identified by expansion of the NSE and $Re$ is the Reynolds number.

The analysis up to this point is very similar to the development of the linearized, fourth-order Orr-Sommerfeld-Squire operator of linear stability theory, with the exceptions that the $(k,n,\omega)=(0,0,0)$ mode is identified as the turbulent mean velocity rather than the laminar solution, and the nonlinear forcing terms are explicitly retained in the present analysis.
As such, receptivity concepts relevant to the study of neutrally-stable disturbances in inviscid, linearized laminar flow can be extended to the turbulent case, with the understanding that in the latter case the waves are lightly damped and would asymptotically decay in the absence of forcing $\mathbf{f}$. The form of the resolvent dictates that a nearly-singular, essentially inviscid response occurs at the \emph{critical layer}, where the local mean velocity $U$ is equal to the streamwise convective velocity of the wave.

Given the very selective receptivity of the flow, we may hypothesise that the velocity field is dominated by the largest possible response to forcing, at any given wavenumber and frequency set. Mathematically, we may find and order these orthogonal \emph{response modes} by use of the \emph{Schmidt decomposition} \cite{Young-Hilbert} of the resolvent, which provides, for a given forcing magnitude, the largest possible velocity field response, the second largest and so on.\footnote{The calculations are performed using a modified version of the approach of \cite{Meseguer03} and the experimental mean velocity data of \cite{mckeonmean04}.} In practice, we find that the velocity field given by the first of these response modes is associated with a response typically one to many orders of magnitude larger than that of the second mode. The velocity field observed in a real flow will therefore be well described by the first response mode if the forcing contains a non-negligible component of the correct shape. This assumption of selective receptivity is congruent with an assumption that the resolvent may be approximated by a low rank operator (which has been postulated in various forms by many previous researchers).

Under the further assumption of forcing that is small relative to the magnitude of the response modes, a standard asymptotic analysis \cite{DrazinReid} can be used to describe the scaling of the two regions where viscous effects are required: firstly at the critical layer (to regularize the singularity of the inviscid problem) and at the wall (to meet the wall boundary condition). We call a response mode in which viscous modification at the wall dominates, a \emph{wall} or \emph{attached}  mode, and we call a response mode where viscous modifications at the critical layer dominate a \emph{critical} mode.

In contrast to calculating statistics at a single wall-normal location ($y$) at a given Fourier wavenumber, the response modes provide a basis that predicts the wall-normal amplitude and phase variation of the velocity field. To date, we have examined the structure of the response modes up to a Reynolds number of $Re=2\times 10^4$, of the same order of magnitude as conditions in, for example, a transcontinental natural gas pipeline \cite{McKeonSci10}.  Here we present key results at $Re=2 \times 10^3$, a condition that is achievable in both experiment and state-of-the-art simulation.

% FIGURE 1

% See structure-figs.tex for figure commands and generation
\begin{figure}
\centering
\includegraphics[width = \textwidth]{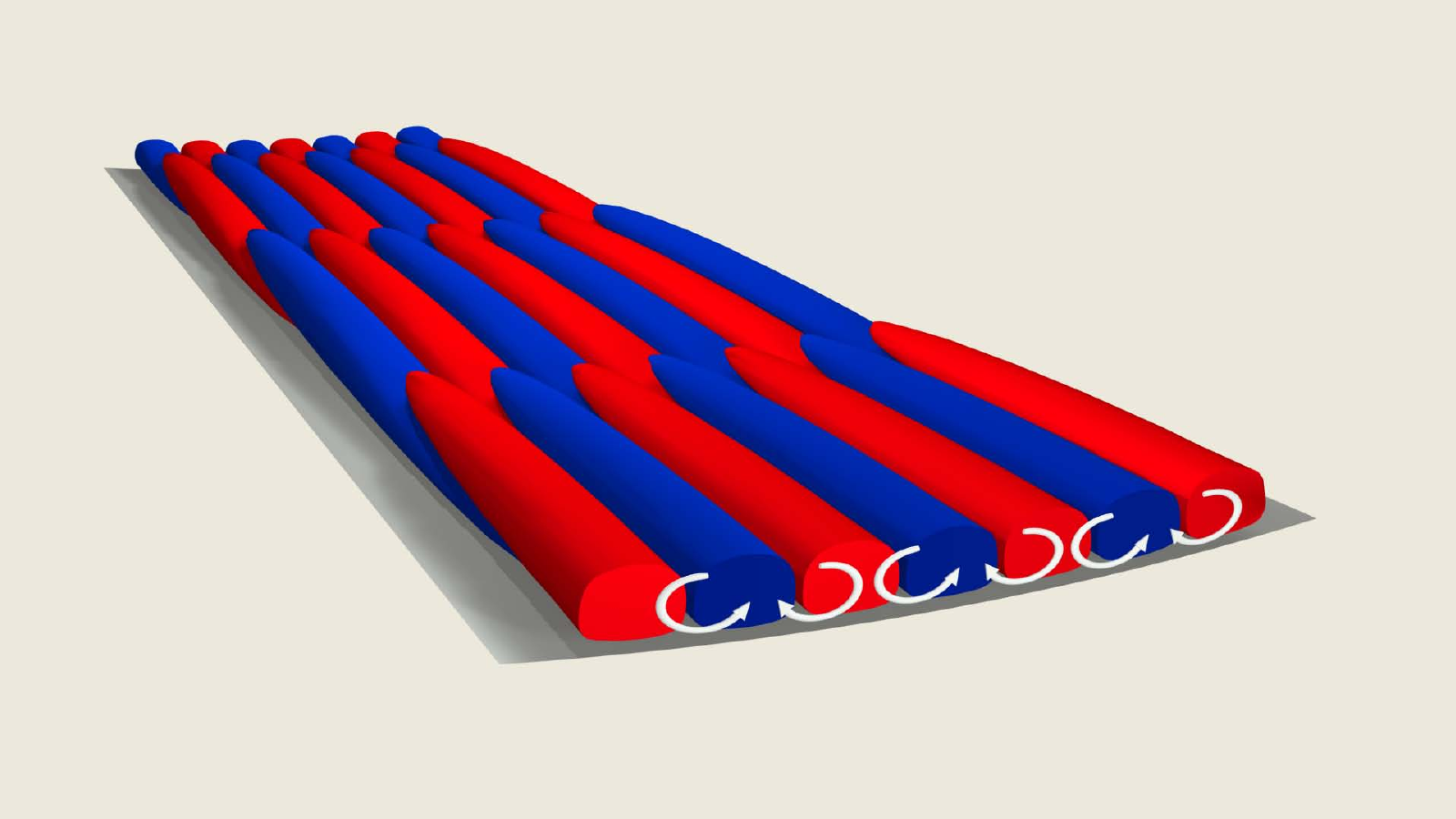}
\caption{Shape of the first singular mode representative of the dominant near wall motions, $(\lambda_x^+,\lambda_z^+,U_x^+)=(1000,100,10)$. Color denotes isosurfaces of streamwise velocity (streaks), where red and blue correspond to high and low velocity respectively relative to the mean flow (heading into the page), and the arrows show the sense of the in-plane velocity field.}
%Prediction of scaling of near-wall peak in u and demonstration of mode shape. I.e. essentially JFM figure 10a followed by some 3D representation of near-wall cycle.  Isocontours of u with v indicated by arrows?
\label{fig:near wall motions}
\end{figure}

Figure \ref{fig:near wall motions} shows the streamwise velocity component arising from a combination of the left and right going helical velocity response modes with a wavenumber-frequency combination representative of the dominant motion near the wall, namely streamwise and spanwise wavelengths of a thousand and one hundred viscous units, respectively, and a convection velocity of ten times the friction velocity, $(\lambda_x^+,\lambda_z^+,U_x^+)=(1000,100,10)$.  The distinctive pattern of rolling motions aligned in the streamwise direction and strong, alternating inclined streaks of fast and slow streamwise velocity $u$ shown in Figure \ref{fig:near wall motions} is entirely consistent with visual and quantitative observations of the near-wall region in canonical flows~\cite{Schoppa02}.  In McKeon \& Sharma \cite{McKeon2010}, we showed that the wall-normal location of the peak intensity of $u$ associated with this mode is independent of Reynolds number. This scaling result is borne out by experimental measurements over a range of Reynolds numbers.

% FIGURE 2

\begin{figure}
\centering
\includegraphics[width = \textwidth]{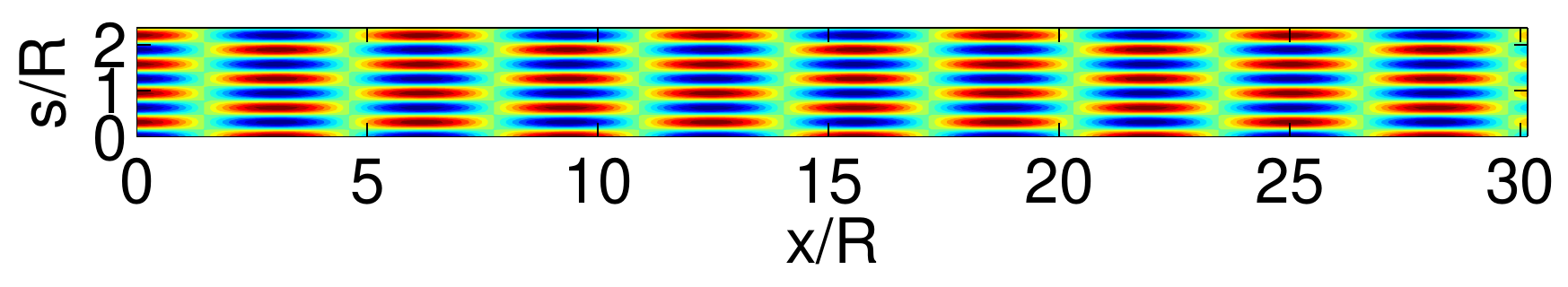}
\includegraphics[width = \textwidth]{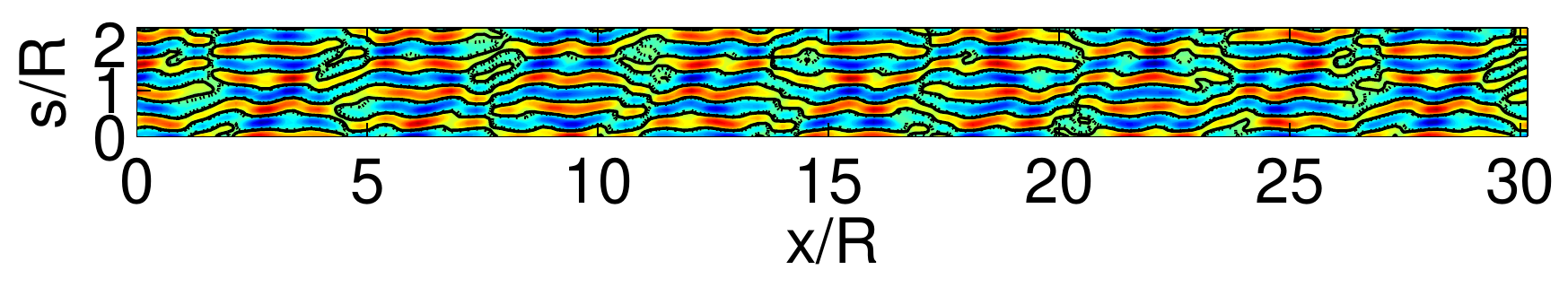}
\includegraphics[width = \textwidth]{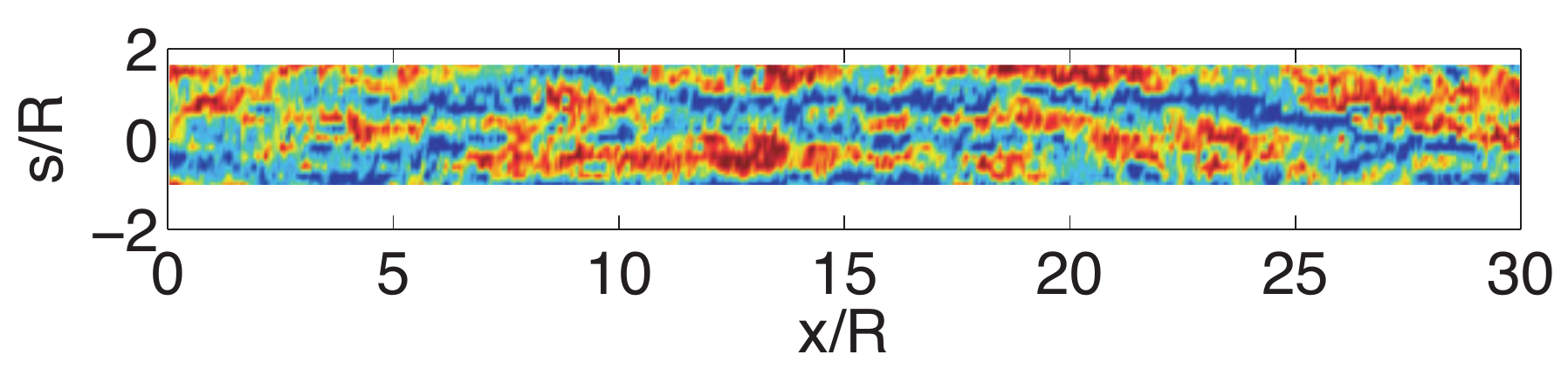}
\caption{Regeneration of apparent ``meandering'' of the VLSMs. The top panel contains isosurfaces of the streamwise velocity at $y/R = 0.15$ for the summation of left- and right-going VLSM modes and the middle panel shows how even longer coherence, of the order of the panel length, can be obtained by superposing the velocity response mode pair fields from two additional (shorter) modes with $(k,n,\omega) = (4.7,\pm12,0.2)$ and $(6.2,\pm15,0.6)$ with amplitude $75\%$ of the VLSM modes. The bottom panel shows the experimental results of Monty \etal~\cite{Monty07}.}
%Prediction of VLSM peak scaling and explanation of ``meandering". I.e. JFM figure 13, but in non-premultiplied form ($y^+ vs R^+$) and my APS09 talk, slide 11, bottom left panel, together with similar plot for overlay of several modes?}
\label{fig:VLSM meandering}
\end{figure}

Our framework can also be used predictively to describe the Reynolds number dependence of the location of peak VLSM energy, $y^+_{pk}$, and to understand the origin of VLSMs.  A mode that is both attached to the wall and critical has a special significance. In pipe flow, for parameters representative of a VLSM, $(k,n)=(1,10)$ approximately, this condition occurs when the convective velocity is $2/3$ of the centerline velocity, independent of Reynolds number, leading to the prediction
\[y^+_{pk} = 0.8 Re^{2/3}.\]
The agreement between this theoretically derived expression and the experimental results of~\cite{Morrison04} reported by McKeon \cite{mckeonAIAA08} is remarkably good \cite{McKeon2010}. This prediction differs from the $Re^{1/2}$ law appropriate for a boundary layer.

Having described important statistical aspects of pipe flow, we now turn our attention to structural considerations.
A simple superposition of first response modes at different wavenumber-frequency combinations also sheds light on the controversy surrounding the true length of the VLSM motions.  Even the addition of the velocity fields corresponding to only two additional pairs of response modes with similar amplitudes to the VLSM mode described above quickly leads to the observation of apparently meandering structures with length far greater than six radii, as shown in Figure \ref{fig:VLSM meandering}. The visual similarity is striking to experimental data shown in the lowest panel of the figure (from \cite{Monty07}). The meandering phenomenon is revealed to be an artefact of the many response modes that are present in a real flow combining with the energetic content of the VLSM itself, effectively decorrelating the VLSM mode.

% FIGURES 3, 4

\begin{figure}
\centering
\includegraphics[width = \textwidth]{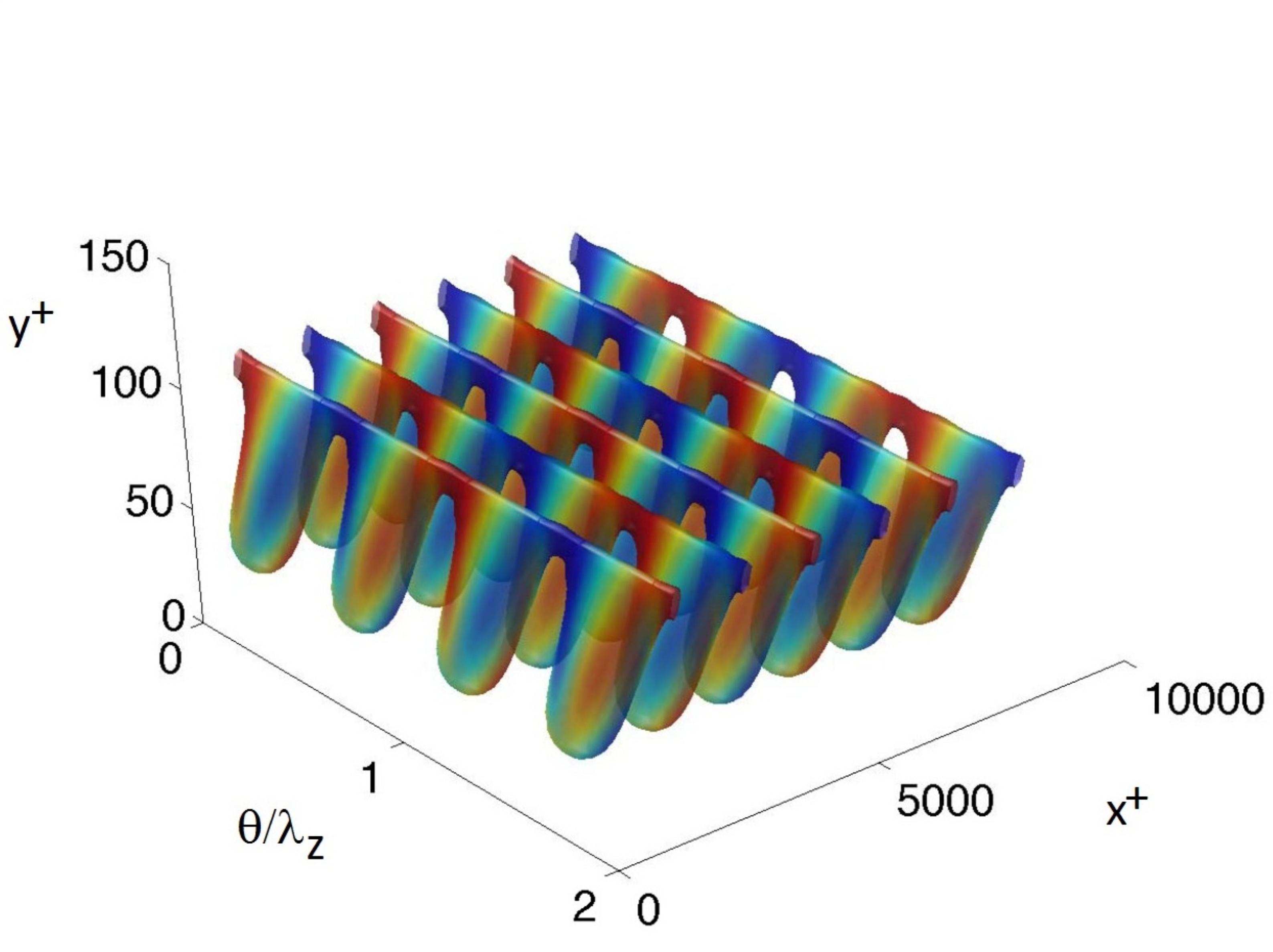}
\includegraphics[width = \textwidth]{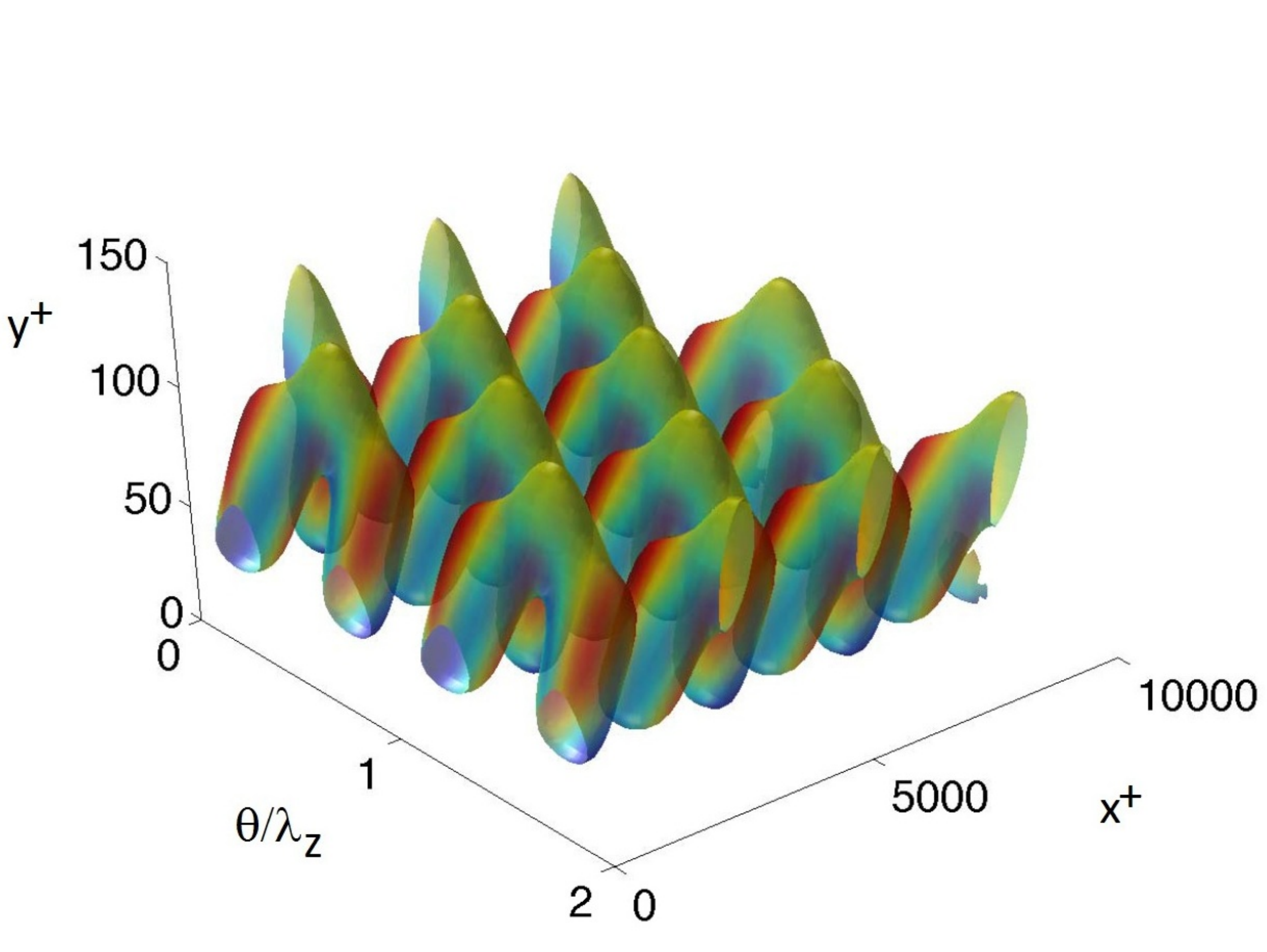}
\caption{Isosurfaces of constant swirling strength for the $(k,n,\omega)=(4.5,\pm10,1.67)$ velocity response mode (three wavelengths are shown in the streamwise and two in the spanwise directions) at $Re=1800$, color-coded with the sense of the azimuthal rotation. Blue and red denote pro- and retro-grade swirl (or rotation in and counter to the sense of the classical hairpin vortex), respectively. Top: under a Galilean transformation (i.e. constant convection velocity subtracted throughout the field of view) there are even numbers of prograde and retrograde vortices. Bottom: with the mean velocity profile added, the retrograde vortices disappear and the prograde ones are strengthened.}
%Panels of velocity isocontours and 3D swirl field for representative wavenumber-frequency combination, no mean field. Ideally panels of $u,v,w$ in $x-y$ plane then fourth panel showing 3D swirl field.  Too many panels?}
\label{fig:hairpin isosurfaces}
\end{figure}

\begin{figure}
\centering
\includegraphics[width = \textwidth]{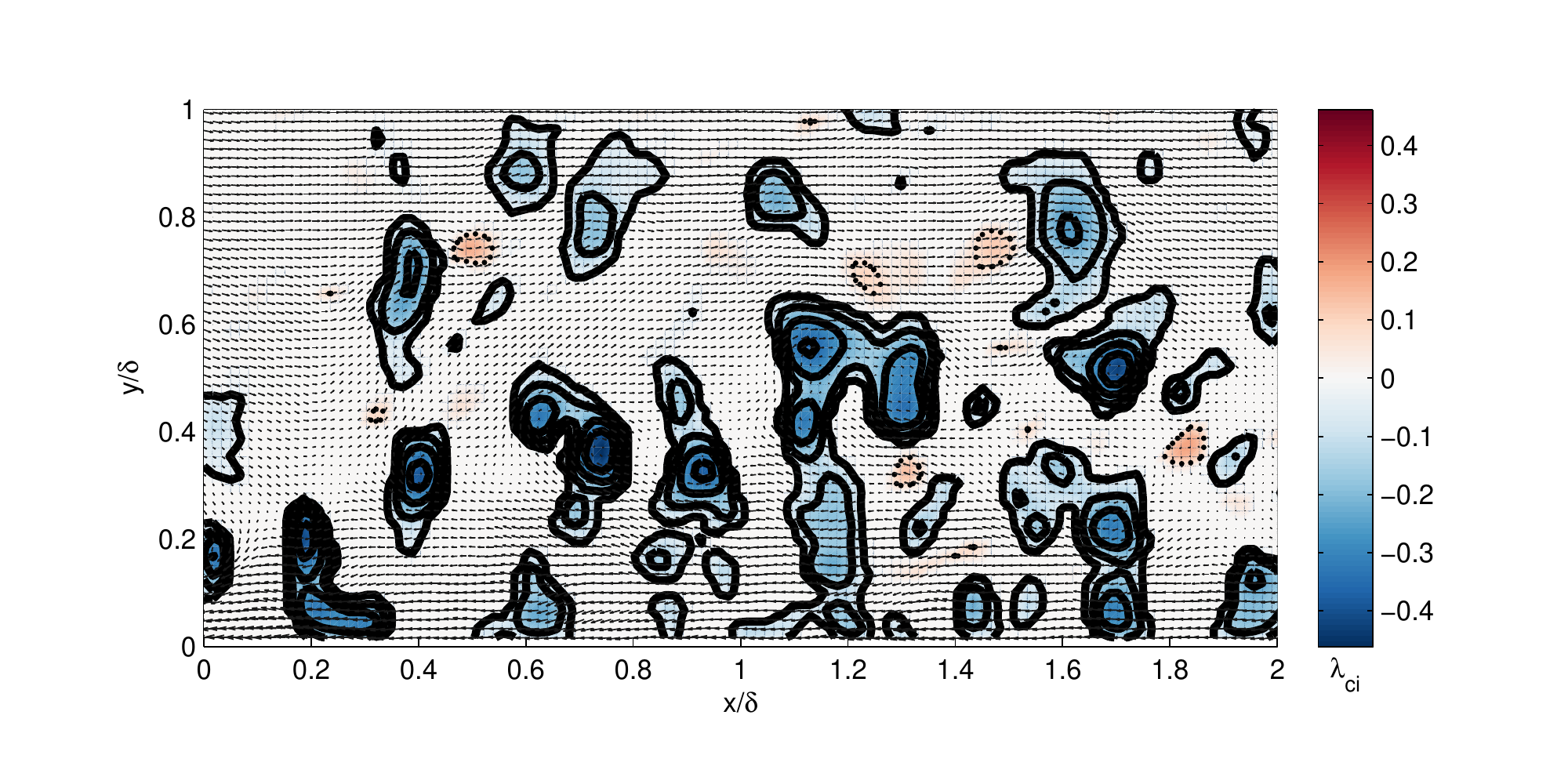}
\includegraphics[width = \textwidth]{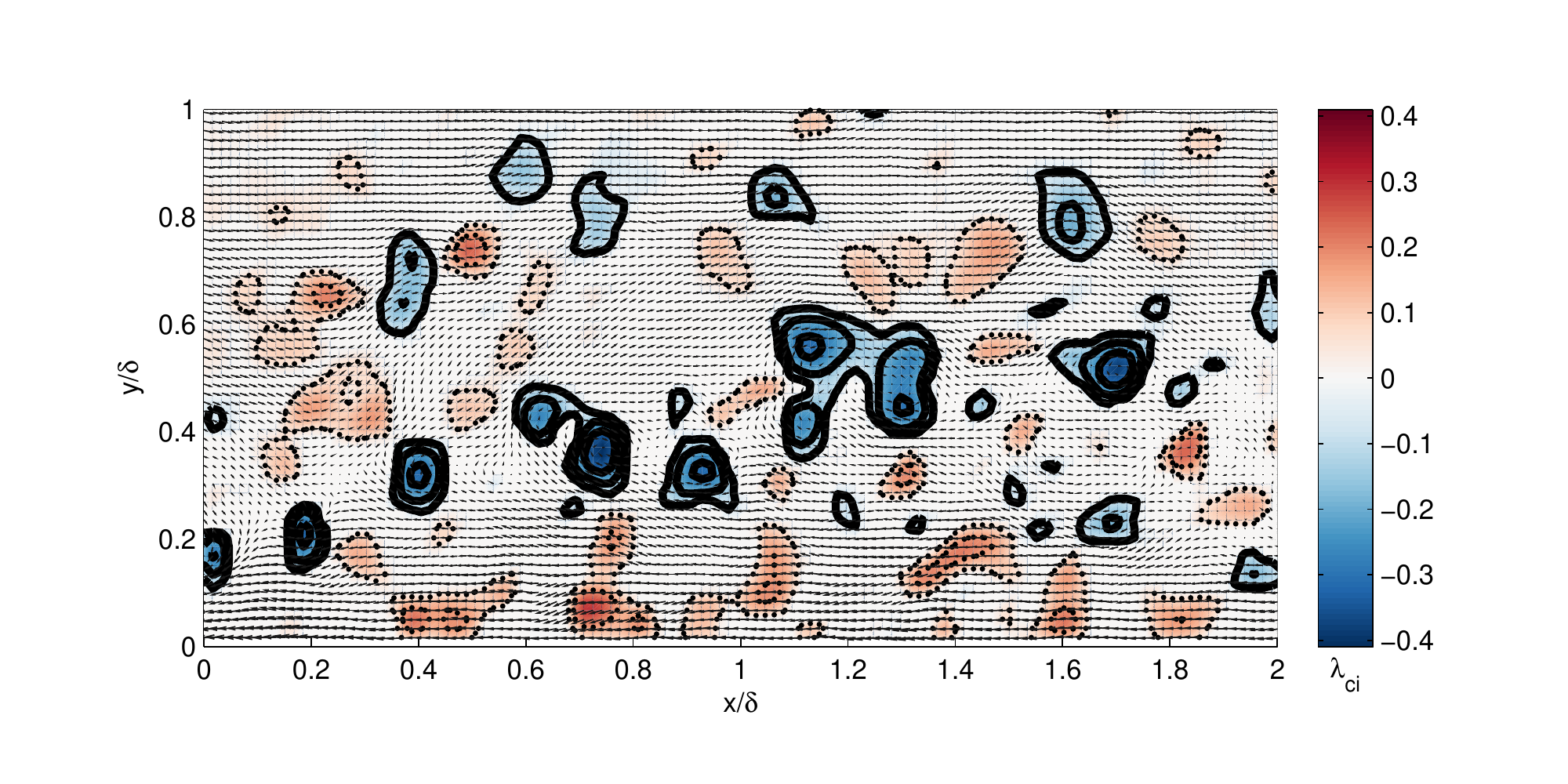}
\caption{Experimental streamwise and wall-normal velocity fields obtained using PIV in a zero pressure gradient turbulent boundary layer at a similar Reynolds number to the results of Figure~\ref{fig:hairpin isosurfaces}. Top: Galilean transformation (i.e. constant convection velocity subtracted throughout the field of view).  Bottom: Reynolds decomposition (i.e. local mean velocity subtracted at each wall-normal location).  Blue and red denote pro- and retro-grade swirling strength, respectively, that exceeds the chosen threshold magnitude and the swirl is overlaid on a quiver plot of the local in-plane velocities.}
%2D swirl field overlaid on quiver plot of $(u,v)$ with mean shear, then cartoons of true swirl field and observed field.}
\label{fig:hairpin PIV}
\end{figure}

The three-dimensional velocity field associated with the attached near-wall mode shown in Figure \ref{fig:near wall motions} gives an intuitive hint as to the locations of coherent vorticity associated with this type of mode. We identify structure through measures which distinguish between the shear and rotational components of vorticity, namely the symmetric and anti-symmetric components of the velocity gradient tensor, $\nabla \mathbf{v}$.  While any of the commonly used measures (see Chakraborty \etal~\cite{Chakraborty05}) give very similar results, we highlight isosurfaces of \emph{swirling strength}\footnote{$\lambda_{ci}$, or the imaginary part of the complex conjugate eigenvalue pair associated with the velocity gradient tensor} for a wall mode with $(k,n,\omega)=(4.5,\pm 10,1.67)$ in Figure \ref{fig:hairpin isosurfaces}a. The surfaces are colored by the magnitude of the azimuthal vorticity, with blue corresponding to \emph{prograde} vortices, with the same sense of rotation as the traditional hairpin, and red to \emph{retrograde} vortices, with the opposite sense of rotation. The latter have been reported in the literature to occur relatively infrequently \cite{Natrajan07,CS05}. Our model predicts an equal distribution of prograde and retrograde vortices associated with each wall mode. However, in the presence of a mean velocity profile with decreasing shear in the wall-normal direction, the retrograde vortices are suppressed while the prograde ones are reinforced, leading to a distribution of vortices that is consistent with experimental observations, Figure \ref{fig:hairpin isosurfaces}b. In this figure, the mean shear is sufficiently strong to completely suppress the retrograde vortices below the swirl threshold selected for plotting. We emphasize that the swirl field regenerated from the full range of modes projected out, for example, from a direct numerical simulation would reproduce the full swirl field: we explore here a decomposition of the swirl field around the mean shear, which precludes consideration of the contribution of hairpin heads to the mean shear itself, as explored by other authors, e.g. \cite{Adrian00}.  There are two points to note here, firstly, this is a simple manifestation of the mean shear being the only net source of azimuthal vorticity, and secondly, swirling strength is not a distributive operation. In other words, $\mathrm{swirl}(a+b)\neq \mathrm{swirl}(a)+\mathrm{swirl}(b)$. Thus the phenomenon is a direct consequence of the diagnostics commonly used to identify structure. To illustrate, consider the example of swirling strength in a two-dimensional, streamwise/wall-normal velocity field $(u,v)$. Examination of the expression for $\lambda_{ci}$ given in the equation below shows that the mean shear will lead to a reduction in apparent swirling strength in regions where $\partial v/\partial x >0$, as associated with retrograde vortices.
\[\lambda_{ci} = \frac{1}{2}\mathrm{Im} \sqrt{\left(\frac{\partial u}{\partial x} + \frac{\partial v}{\partial y}\right)^2 - 4\left(\frac{\partial u}{\partial x}\frac{\partial v}{\partial y}-\frac{\partial u}{\partial y} \frac{\partial v}{\partial x}\right)}.\]
%....~~~\mathrm{OR}~~~Q=...
%Clearly the situation is further complicated in the presence of multiple modes, all contributing local velocity gradients in a superposition.

Experimental velocity fields obtained in a zero pressure gradient turbulent boundary layer using particle image velocimetry (PIV) %\cite{Jacobivort10}
confirm this observation. Figure \ref{fig:hairpin PIV}a shows the more usual Galilean decomposition of the data (where a constant convection velocity has been subtracted from all locations in the field). There is a prevalence of prograde vortices, denoted in blue, whereas the Reynolds-decomposed field of Figure \ref{fig:hairpin PIV}b, in which the local mean velocity has been subtracted from all wall-normal locations, shows an essentially equal number of prograde and retrograde cores. The positive local $\partial u/\partial y$ associated with the most energetic mode in each picture explains the apparent alignment of hairpins into hairpin packets identified with ramp-like shear layers, while the relative motion between fast- and slow-moving modes explains the apparent wall-normal growth and decay of individual vortices, and renders the concept of relative phase between modes a moot point. The linearity of our model reveals the crucial importance of a swirl diagnostic that is not affected by mean shear, a failing that has impeded our understanding of the appropriate distribution of coherent vortical structures that corresponds to the appropriate mathematical understanding.

It is clear that a very complex velocity field can be obtained simply by superposing modes with different $(k,n,\omega)$ (and therefore different convective velocities) and amplitudes, and with associated local velocity gradients. Critically, because swirling strength is not a distributive operation, the swirl field cannot be simply determined from a superposition of the swirling strength associated with the individual response modes.

% Why the main result is significant
Understanding turbulence has proven difficult because it requires modeling at all scales.  Our approach predicts and explains aspects of the vortical structure and velocity statistics associated with turbulent flows that previously have been explained only in phenomenological terms and have been identified only in experiments or computationally costly direct numerical simulation, which is limited to Reynold numbers much lower that those relevant to most applications. The linearity of the processes driving wall turbulence permits superposition of the response modes, a surprising result given the understanding of the importance of nonlinear dynamics in wall turbulence, although that linear processes are also important is well known. This linearity is revealed by formulating the NSE as a forcing-response problem. We have studied pipe flow in the current work, but the same approach can be applied to both internal and external flows with simple modifications. The results shown in this paper were generated in seconds using a standard laptop computer, and the approach extends to higher Reynolds numbers. The only limitation is the numerical precision required to deal with the near-singular response to the most amplified forcing.

% Broader perspective for a general audience
A fundamental understanding of the underlying structure of wall turbulence has eluded researchers for decades. Excitingly, our results close in on a reconciliation of the statistical and structural interpretations of such flows by working from the NSE and an assumed mean profile.
The understanding of the different types of mode and where they occur in the flow has significant implications for the prediction of Reynolds number trends and the modeling of turbulent activity at reduced computational cost.
While the forcing in our model is currently unstructured, such that we do not determine the appropriate amplitudes at individual modes, the possibility to ``close the loop" and formulate a reduced order model of turbulent pipe flow is compelling: under a self-consistent combination of modes the assumed mean velocity profile will be generated, such that a self-sustaining system can be designed. The potential to design rigorous control techniques for such a system with the objective of enforcing favorable turbulence characteristics is a natural, and plausible, next step.

\section*{Acknowledgements}
 The authors acknowledge financial support from the Air Force Office of Scientific Research grants FA9550-08-1-0049 and FA9550-09-1-0701, Program Manager John Schmisseur (B.J.M. \& I.J.) and an Imperial College Junior Research Fellowship (A.S.).

Correspondence and requests for materials
should be addressed to B.J.M.~(email: mckeon@caltech.edu) or  A.S.~(email: ati@ic.ac.uk).

\bibliography{refs_hairpins}

\end{document}